\documentclass[aps,prl,twocolumn,nopacs,floats,superscriptaddress]{revtex4-1}
\usepackage{graphicx}
\usepackage{dcolumn}
\usepackage{bm}
\usepackage{amsmath}
\usepackage{color}
\usepackage{amssymb}
\usepackage{ulem}
\usepackage{xcolor}
\usepackage{subfigure}

\definecolor{cardinal}{rgb}{0.6,0,0}
\definecolor{darkgreen}{rgb}{0,0.4,0}
\definecolor{golden}{rgb}{0.92, 0.7, 0}
\definecolor{midnight}{rgb}{0, 0, 0.5}
\definecolor{darkblue}{rgb}{0, 0, 0.7}

\def\he4{$^4$He}
\def\hel3{$^3$He}
\def\Am3{\AA$^{-3}$}
\def\beq{\begin{equation}}
\def\eeq{\end{equation}}

\newcommand{\be}{\begin{equation}}
\newcommand{\ee}{\end{equation}}
\newcommand{\bea}{\begin{eqnarray}}
\newcommand{\eea}{\end{eqnarray}}
\newcommand{\bse}{\begin{subequations}}
\newcommand{\ese}{\end{subequations}}

\begin{document}
%%%%%%%%%%%%%%%%%%%%%%%%%%%%%%%%%%%%%% AUTHORS %%%%%%%%%%%%%%%%%%%%%%%%%

\author{Anatoly Kuklov}
\affiliation{Department of Physics \& Astronomy, College of Staten Island and the Graduate Center of
CUNY, Staten Island, NY 10314}

\author{ Lode  Pollet}
\affiliation{Department of Physics and Arnold Sommerfeld Center for Theoretical Physics, Ludwig-Maximilians-Universit\"at M\"unchen, Theresienstr.~37, M\"unchen D-80333, Germany}
\affiliation{Munich Center for Quantum Science and Technology (MCQST), Schellingstr.~4, D-80799 M\"unchen, Germany}
%\affiliation{Arnold Sommerfeld Center for Theoretical Physics, Ludwig-Maximilians Universit\"at, Theresienstrasse 37, 80333 M\"unchen, Germany}

\author{Nikolay Prokof'ev}
\affiliation{Department of Physics, University of Massachusetts, Amherst, MA 01003, USA}

\author{Leo Radzihovsky}
\affiliation{ Department of Physics and Center for Theory of Quantum Matter, University of Colorado, Boulder, CO 80309}

\author{Boris Svistunov}
\affiliation{Department of Physics, University of Massachusetts, Amherst, MA 01003, USA}
\affiliation{Wilczek Quantum Center, School of Physics and Astronomy and T. D. Lee Institute, Shanghai Jiao Tong University, Shanghai 200240, China}

%%%%%%%%%%%%%%%%%%%%%%%%%%%%%%%%%%%%%%%%%%%%%%%%%%%%%%%%%%%%%%%%%%%%%%%%%%%%%%
\title{Universal Correlations as Fingerprints of Transverse Quantum Fluids }
%%%%%%%%%%%%%%%%%%%%%%%%%%%%%%%%%%%%%%%%%%%%%%%%%%%%%%%%%%%%%%%%%%%%%%%%%%%%%%
\begin{abstract}
  We study universal off-diagonal correlations in transverse quantum
  fluids (TQF)---a new class of quasi-one-dimensional superfluids
  featuring long-range-ordered ground states.  These exhibit unique
  self-similar space-time relations scaling with $x^2/D\tau$ that serve
  as fingerprints of the specific states.  The results obtained
  with the effective field theory are found to be in perfect agreement
  with {\it ab initio} simulations of hard-core bosons on a lattice---a simple microscopic realization of TQF.
  This allows an accurate determination---at nonzero temperature and finite system size---of
  such key ground-state properties as the condensate and superfluid
  densities, and characteristic parameter $D$.
\end{abstract}

\maketitle

{\it Introduction.}
The concept of a transverse quantum fluid (TQF) originally emerged in the context of
superfluid edge dislocations in $^4$He \cite{Kuklov2022,TQF1} in an attempt to
explain the observed ``flow-through-solid'' phenomena \cite{Hallock,Hallock2012,Hallock2019,Beamish,Moses,Moses2019,Moses2020,Moses2021}.
It was subsequently realized \cite{TQF2} that there exists a
broad class of quasi-one-dimensional superfluids featuring similar
properties. Examples include a superfluid edge of
a self-bound droplet of hard-core bosons on a two-dimensional (2D)
lattice, a Bloch domain wall in an easy-axis ferromagnet, and a phase
separated state of two-component bosonic Mott insulators with the
boundary in the counter-superfluid phase (or in a phase of
two-component superfluid) on a 2D lattice. The TQF
state is a striking demonstration of the conditional
character of many dogmas associated with superfluidity and its order parameter field,
such as elementary excitations, in general,  and the ones obeying the Landau criterion in particular.
In sharp contrast with Luttinger liquids---the standard paradigm for one-dimensional quasi-long-ranged
superfluids---TQFs feature long-range-ordered ground states supporting persistent currents.

\begin{figure}[!htb]
%\vskip-8mm
\includegraphics[width=0.9 \columnwidth]{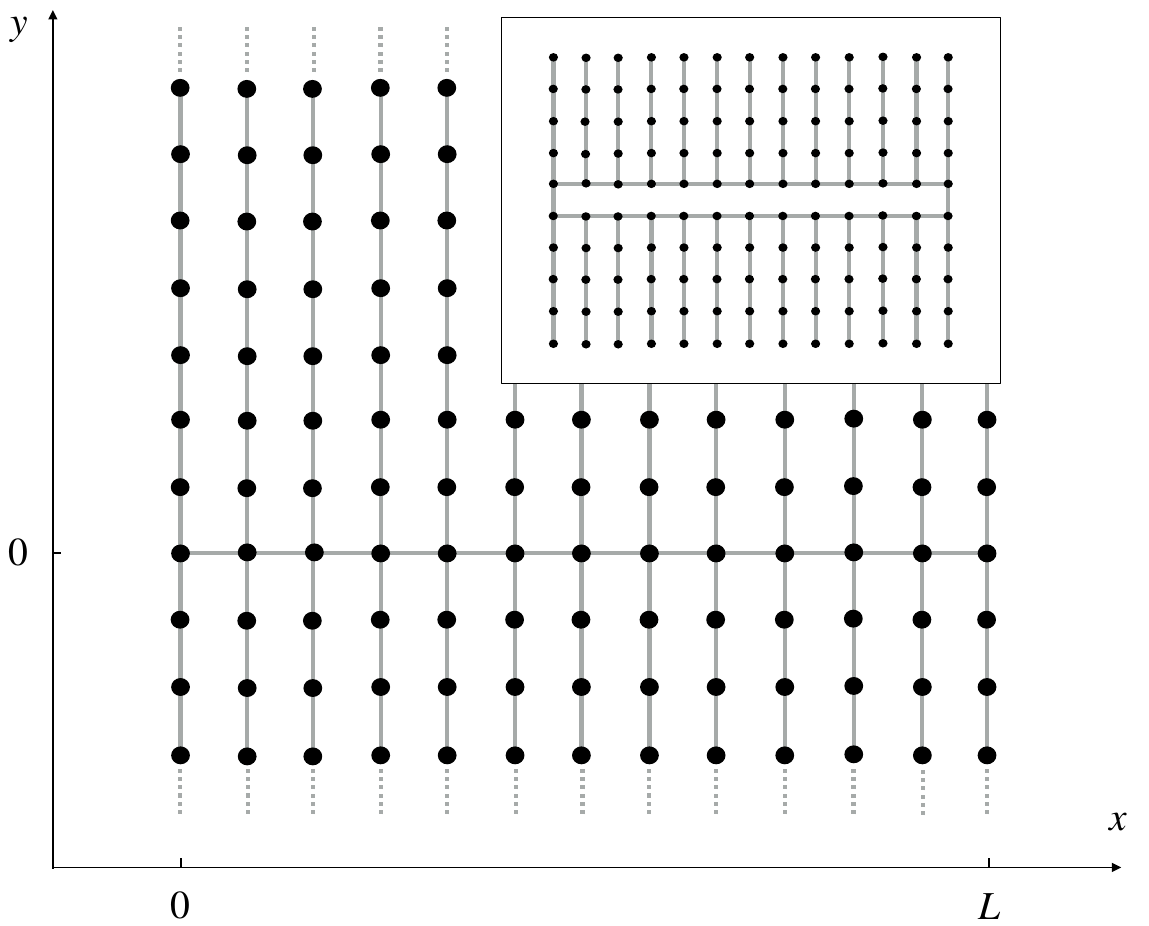}
%	\vskip-8mm
\caption{Minimal iTQF model: Hard-core bosons
at filling factor $n=0.5$ on the square lattice, with
all hopping amplitudes between sites in
the $x$-direction suppressed, except along the single row at $y=0$.
The non-zero hopping amplitudes are marked by solid lines. Inset: Image of a system of finite length with periodic boundary conditions, 
supporting a superflow along the central loop, which we expect to be implementable experimentally. }
\label{iTQF_model}
\end{figure}

Currently, there are two known classes of TQFs \cite{TQF2}:
(i) systems featuring well-defined elementary excitations with a quadratic dispersion, $\omega = D k^2$, and
(ii) so-called {\it incoherent} TQF systems (iTQFs) where the
dynamics of phase fluctuations has diffusive character, $\omega = -i D k^2$,
i.e., they lack elementary excitations.
All the examples mentioned above belong to the class (i).
A minimal iTQF model, most relevant for quantum emulation with
ultracold atoms and efficient {\it ab initio} numeric simulation, is
presented in Fig.~\ref{iTQF_model}. The system consists of hard-core 
bosons hopping on a lattice, Josephson-coupled only via a single 1D path
(horizontal links at $y=0$), which ultimately forms the iTQF channel. 

%For computational convenience (that can be easily generalized) we focus on the
For computational convenience we focus on systems with particle hole symmetry at half filling
%particle-hole symmetric filling factor of $1/2$ 
and with the same hopping amplitude $t$ between all connected sites. 
Working in energy units of $t$ and length
scales of the lattice spacing, our microscopic model is parameter-free.
The ground-state condensate density, $n_0$, superfluid stiffness, $n_s$,
and the diffusion constant, $D$, are the key quantities uniquely
characterizing the universal iTQF properties of the system.

%%%%%%%%%%%%%%%%%%%%%%%%%%%%%%%%%%%%%%%%%%%%%%%%%%
%%%%%%%%%%%%%%%%%%%%%%%%%%%%%%%%%%%%%%%%%%%%%%%%%%%%%%%

\begin{figure}[!htb]
%\vskip-8mm
\includegraphics[width=0.9 \columnwidth]{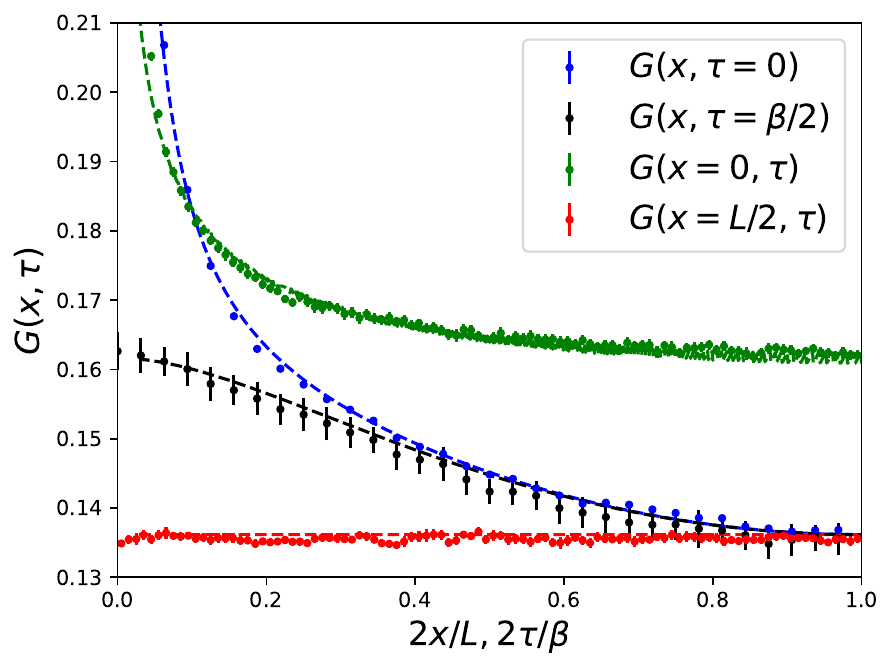}
%	\vskip-8mm
\caption{Green's functions for the model Eq.~(\ref{eq:Zi}). The dashed lines are the fits by the iTQF theory. The fit parameters have been optimized for $G(x, \tau=0)$ and are then reused for the other cases. The superfluid stiffness $n_s =0.5279(4)$ is obtained from the fluctuations in winding numbers along the $x$-axis, $n_s = (L/\beta) \langle W_x^2 \rangle$. The fit results are $n_0 = 0.1481(4)$ and $\gamma = 8.86(9)$ (or $D \approx 1.28$).   }
\label{fig:Zi}
\end{figure}

In this Letter, we show---based on {\it ab initio} worm algorithm
quantum Monte Carlo simulations \cite{Worm} and predictions of the effective field theory---that
finite-size/finite-temperature behavior of the minimal model shown in
Fig.~\ref{iTQF_model} [and also that of physically similar but microscopically different model (\ref{eq:Zi})] is perfectly and in full detail described by the iTQF effective field theory. This is so even for relatively small system
sizes, $L$, and  with temperatures  $T^{-1} \equiv \beta \, \lesssim\,  L$, providing a powerful tool to clearly
reveal the unique off-diagonal correlations---the fingerprint of iTQF---experimentally.
Our key results are presented in Figs.~\ref{G} and \ref{fig:Zi}.

\noindent{\it Effective field theory model:}
We begin with the iTQF Euclidean low-energy effective action \cite{TQF2} ($\hbar=1$ throughout)
\be
S_{\rm iTQF} = { 1 \over 2}\sum_{\omega, k}  \left[K_b|\omega|
  + n_s k^2 \right]|\phi_{\omega,k}|^2 \, ,
\label{theta2}
\ee
where the parameter $K_b$, or, equivalently, the diffusion coefficient $D=n_s/K_b$ controls the
real-time diffusive dynamics $\omega = -  i Dk^2$ of the
superfluid phase, $\phi(x,\tau)$, for the boson field
$\psi \sim e^{i\phi}$.

Neglecting topological defects (instantons),
the state with a harmonic
action is fully characterized by the Green's function
$G(x,\tau) = \langle \psi^*(x,\tau) \psi(0,0)\rangle$,
straightforwardly obtained from Gaussian integrals:
\be G(x,\tau) \, =\,   n_{\rm uv} \, \Phi(x) \, e^{-\tilde{C}(x,\tau)}  \, ,
\label{G}
\ee
where $n_{\rm uv}$ is the ultraviolet-cutoff-dependent quantity,
\bea
\Phi (x) &=& Z^{-1} \sum_{m=-\infty} ^{\infty} e^{-\beta E_m} \, e^{i 2 \pi m x / L } \, ,
\label{Phi} \\
Z&=& \sum_{m=-\infty} ^{\infty} e^{-\beta E_m} \quad  \mbox{with} \quad E_m=\frac{n_s}{2}\frac{(2\pi m)^2}{L}  \, , \qquad
\eea
is the statistical sum of contributions from persistent current states with phase winding numbers $m$
in the system with periodic boundary [otherwise $\Phi (x)=1$], 
and
\bea
\tilde C(x,\tau) \, &=&\, {1\over 2}\,  \langle \, [\, \phi(x,\tau)-\phi(0,0)\, ]^2 \, \rangle \, ,
\label{phase_corr}\\
&\to &\int_{k,\omega}\;
[1-\cos(k x+ \omega \tau)] \,  c(\omega,k) .
\label{Q}
\eea
The intergration $\int_{k,\omega}\;\equiv \int \frac{d\omega dk}{(2\pi)^2}$ is performed with appropriate UV-cutoff.
The last expression is valid at zero temperature in the thermodynamic limit $L\rightarrow\infty$.
(There is no need for a $\tau$-analog of the $\Phi$-factor for TQF because  $\tau$-windings of the phase are macroscopically expensive.)

The iTQF kernel $c(\omega,k)$ is given by
\be
c(\omega,k)\, = \,
{D \over n_s} \, {1\over |\omega| +D k^2} \qquad \qquad \mbox{(iTQF)}.
\label{iTQF}
\ee
 Similarly, the TQF state is characterized by
$c(\omega,k)\, = \, (D^2/n_s )\, k^2 (\omega^2 +D^2 k^4)^{-1}$, with
$D = \sqrt{n_s/\kappa}$, see Ref.~\cite{TQF1}.
At zero temperature, both states exhibit off-diagonal long-range order even in 1D,
corresponding to a saturation of the integral (\ref{Q}) in the limit
of $|\tau|,|x| \rightarrow\infty$, leading to a finite condensate fraction $n_0$ in the superfluid ground state,
\be
n_0 =  n_{\rm uv} \, e^{-\int_{k,\omega}\;c(\omega,k)}.
\label{n_0_iTQF}
\ee
This brings us to the UV-cutoff-independent Bogoliubov relation
\be
G(x,\tau) \, =\,  n_0\, \, \Phi(x) \,  e^{ C(x,\tau)}\, ,
~~\mbox {for} ~~ D|\tau|+ x^2 \to \infty ,
\label{GenBog}
\ee where
$C(x,\tau)\equiv \tilde C(\infty,\infty) -  \tilde C(x,\tau)$.

Simple analysis reveals self-similarity of $C(x,\tau)$:
\bea
C(x,\tau)\,& =& \, {D  \over n_s |x|}\, g\!\left( D|\tau |/x^2 \right) \label{scaling1} \\
 &=& \, {\sqrt{D}  \over n_s \sqrt{|\tau |} }\,  f \big( |x| / \sqrt{D|\tau |} \big) \, .
\label{scaling2}
\eea
Here $g(\sigma)$  and $f(\zeta)$ are dimensionless scaling functions related to each other by the identity
$f(\zeta) \, \equiv \, g(1/\zeta^2)/\zeta$, $\zeta=1/\sqrt{\sigma}$, with 
explicit expression given by  \cite{Abram} 
\bea
g(\sigma) \, &=& \,  \int_{\omega,k}  \,  {e^{ik+i\omega \sigma} \over |\omega| +  k^2}  = {\rm Re}\left[ {(1+i) e^{i\over 4\sigma}\over 2\sqrt{2\pi \sigma}} \, {\rm erfc}\left( {1+i \over 2\sqrt{2\sigma}} \right)\right] \nonumber
\\
&\approx& \pi^{-1}(1-12 \sigma^2 + \ldots ), \quad {\rm for} \quad \sigma \ll 1\, .\qquad \label{g_iTQF_s} 
\eea
In a complementary limit,
\be
f(\zeta) \approx  {1\over 2\sqrt{2 \pi}} \left(1 -  \zeta^2/4 + \ldots \right), \qquad \mbox{for} \quad \zeta \ll 1\, .
\ee

%%%%%%%%%%%%%%%%%%%%%%%%%%%%%%%%%%%%%%%%
\begin{figure*}[htbp]
	\centering
	\includegraphics[width=0.32\linewidth]{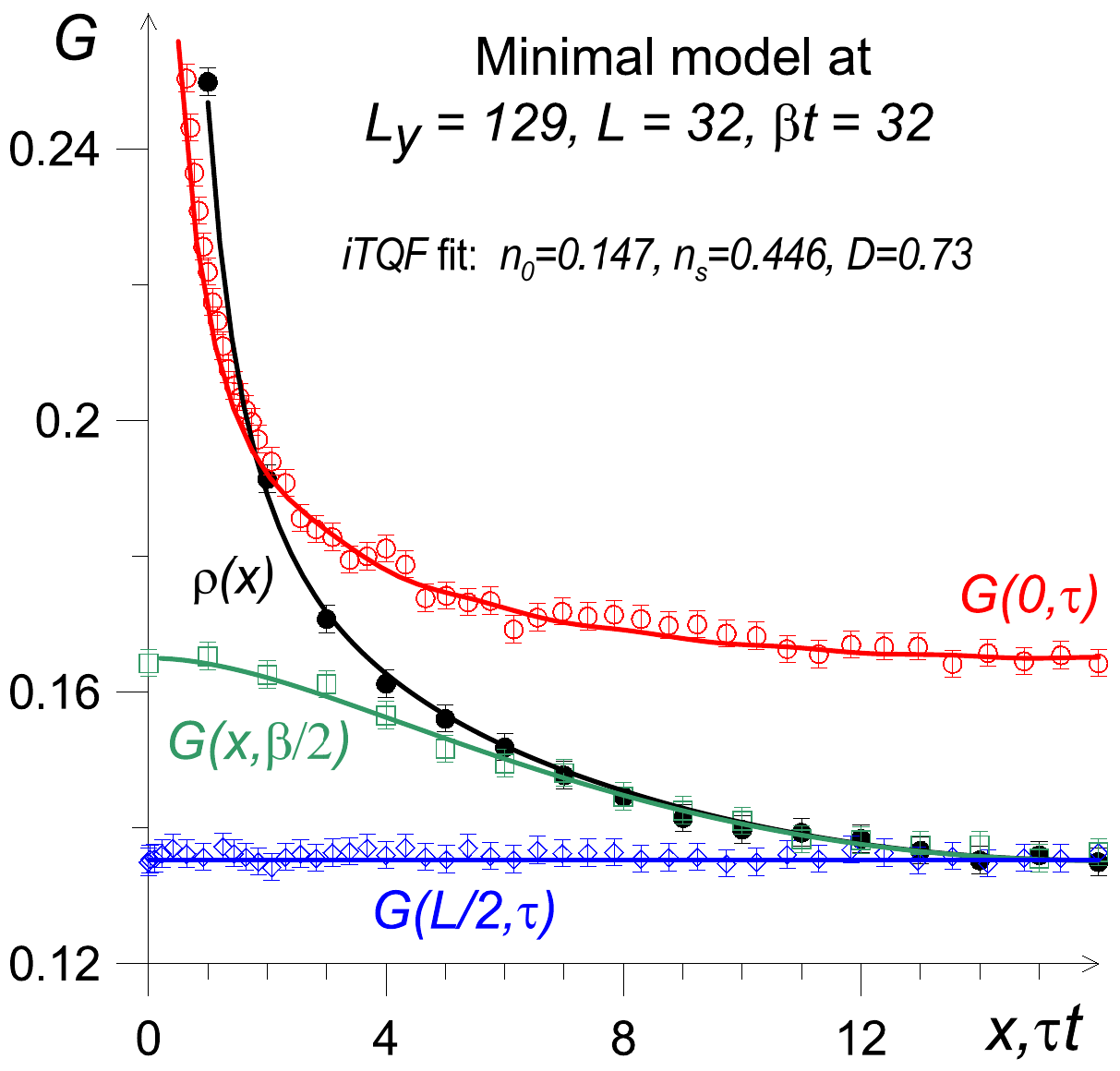}
	\includegraphics[width=0.32\linewidth]{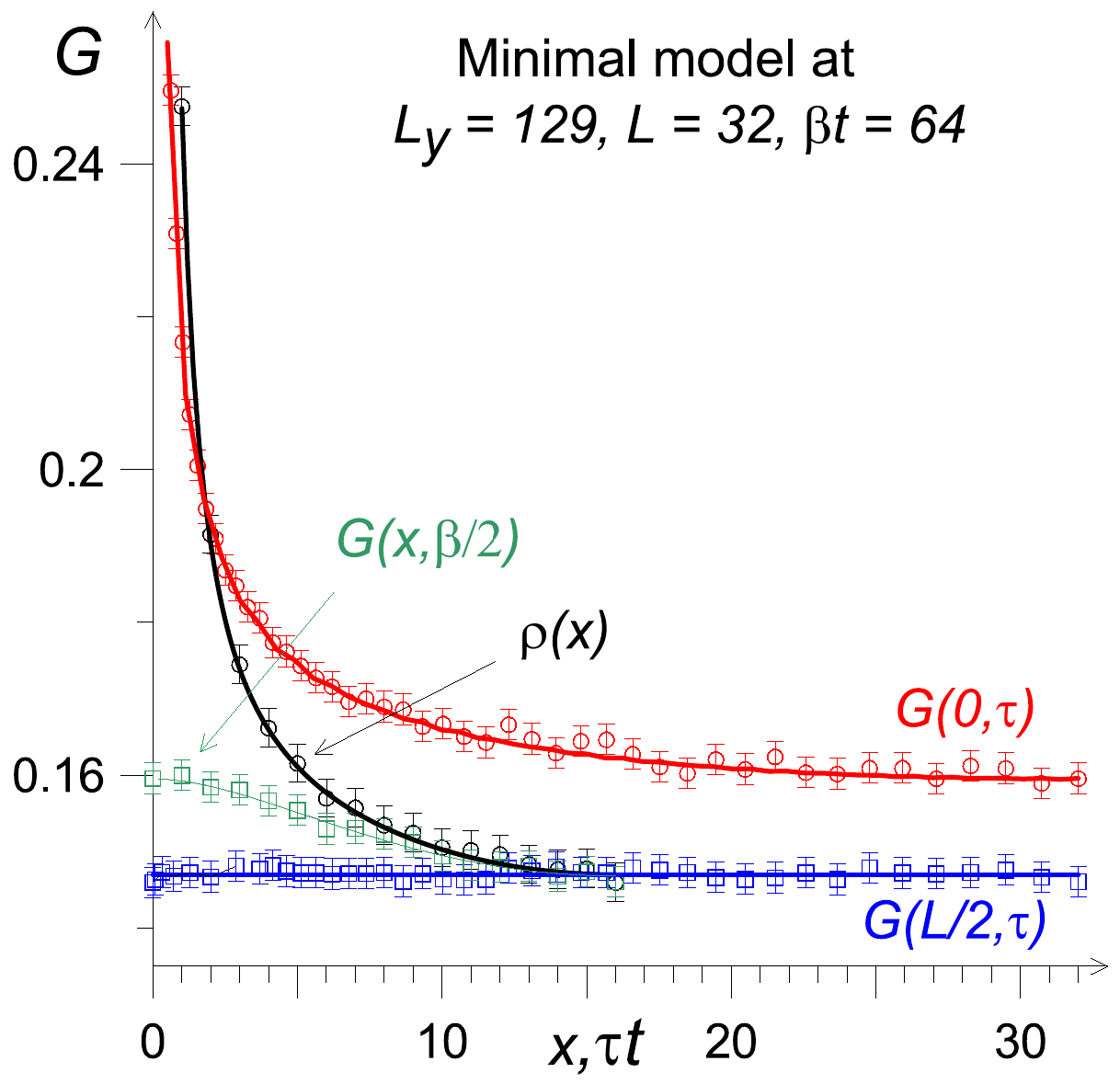}
    \includegraphics[width=0.32\linewidth]{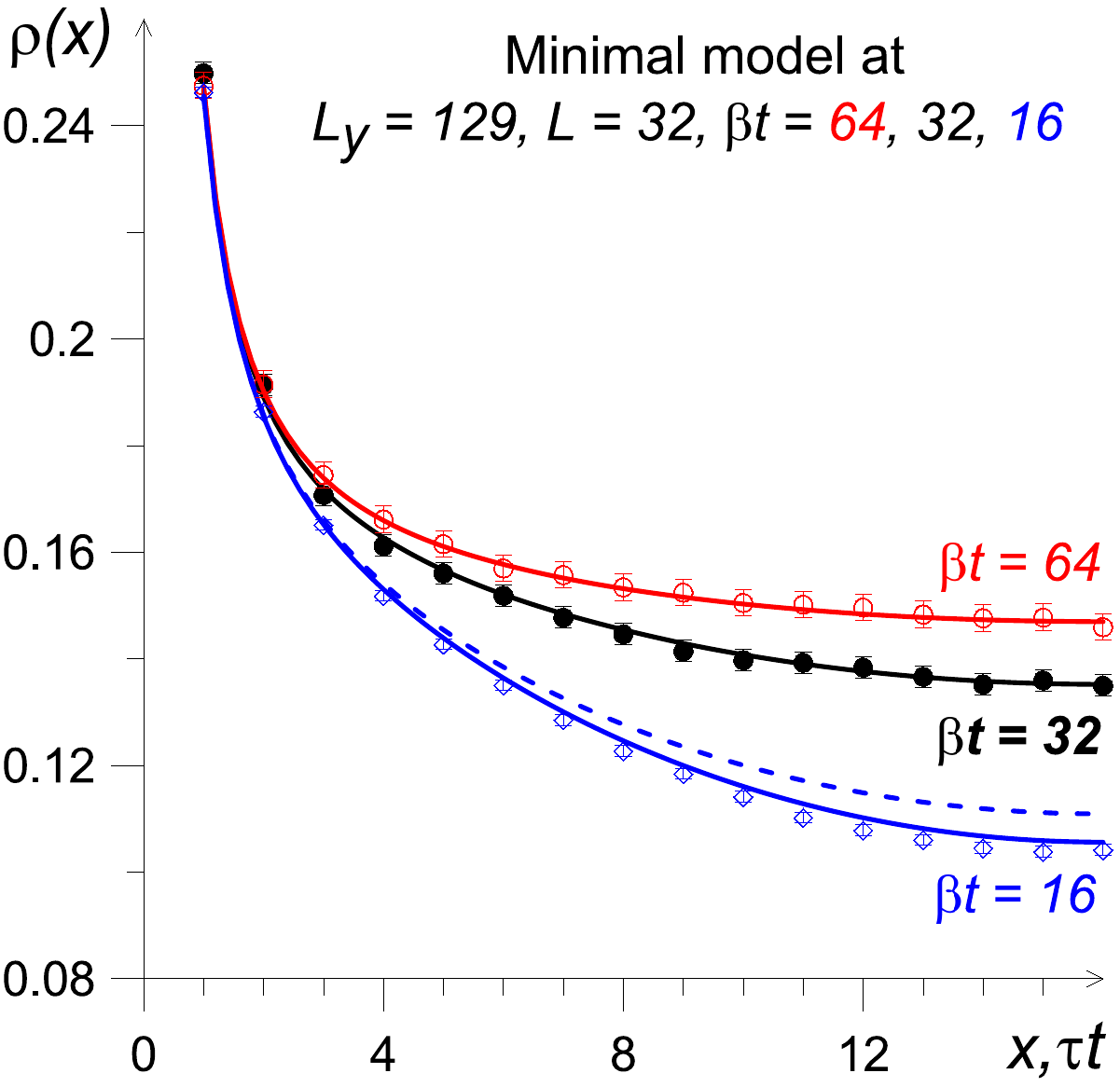}
\caption{\label{G}
Left panel:
Single-particle Green's function $G(x,\tau)$ of the minimal model (Fig.~\ref{iTQF_model})
along four representative space-time directions for $\beta t=32$.
Solid lines are fits to the predictions of the effective iTQF theory using
$n_0$ and $D$ as the only fitting parameters ($n_s=0.446$ is computed from statistics of winding number fluctuations).
The $G(0,0)=0.5$ point is excluded to focus on the universal low-energy behavior.
Central panel:
Green's function for $\beta t=64$ with solid lines representing fit-free predictions
of the effective iTQF theory using parameters determined for $\beta t =32$.
Right panel:
Density matrix $\rho(x)$ for three different temperatures.
Solid lines are theoretical predictions using parameters determined for $\beta t =32$.
In numeric simulations, temperatures are known and the superfluid
stiffness $n_s$ is computed, which leaves only two fitting parameters,
$n_0$ and $D$, to describe all three curves. In the experiment, where not only
$n_s$ but also temperatures may not be known, fitting the curves involves
six parameters. An accurate fit thereby provides not only quantitative
evidence for iTQF state and its ground state properties, but also a protocol
for determining the three temperatures at which $\rho(x)$ were measured.
The dashed line for $\beta t =16$ shows how the fit fails when only the state
with zero phase winding is taken into account, i.e. when $\Phi = 1$.
}
\end{figure*}
%%%%%%%%%%%%%%%%%%%%%%%%%%%%%%%%%%%%%%%

{\it Finite temperature and system size.} For a finite system of
length $L$ with periodic boundary conditions and at a
low but nonzero temperature, the correlator
$C(x,\tau)$ in Eq.~(\ref{GenBog}) is given by (measuring $x$ and
$\tau$ in units of $L/2$ and $\beta/2$, respectively, in all expressions) \be C(x,\tau) =
C_0 + \!\!\!\!
\!\!\!\! \!\!\!\! \!\! \sum_{\scriptsize \begin{array}{c} m,n =-\infty\\
                                           (|m|+|n| \neq
                                           0) \end{array}} ^{\infty} \!\!\!\!
  \!\!\!\! \!\!\!\! e^{i \pi n x} e^{i\pi m\tau} c_{mn} \, ,
\label{Lambda_discr_2}
\ee
where  $c_{mn}\equiv c(\omega_m,k_n) /(\beta L) $, $\omega_m = 2\pi m / \beta$, $k_n  = 2\pi n / L$,
and
\be
C_0 =  \int_{\omega,k}
  c(\omega,k) \; - \!\!\!\! \!\!\!\! \!\! \sum_{\scriptsize \begin{array}{c} m,n \\ (|m|+|n| \neq 0) \end{array}} \!\!\!\!  \!\!\!\! \!\!\ \!\!\!\! \!\!\  c_{mn}  .
\label{Lambda_0}
\ee 
Here, both the integral and the sum have ultraviolet frequency and
momentum cutoffs, which mutually cancel.
%The constant $C_0 $ originates from  our requirement---a matter of convention that we find mathematically natural---that $n_0$ in
The constant $C_0 $ originates from  our convention that $n_0$ in 
(\ref{GenBog}) be exactly equal to the ground-state condensate density in an
infinitely large system. 
[The $m=n=0$ component is absorbed in the definition of condensate;
this will be implied in what follows.]

In the iTQF case, we have
\[
c_{mn} \, = \, {L/\beta\over 4\pi^2 n_s }\,  {1\over n^2 + {\gamma^2 |m| \over \pi^2}} \, ,
\quad \gamma = \sqrt{\pi\over 2}{L\over \sqrt{D\beta}}\quad \mbox{(iTQF)} .
\]
Performing standard summation over $n$, we obtain
\be
C(x,\tau)  = {L/\beta \over 2 n_s} \left[ \frac{x(x-2)}{4} +  \frac{ \lambda_0 + \lambda_1(x,\tau) }{\gamma} \right] \, ,
\label{two_lambdas}
\ee
where
\be
\lambda_1(x,\tau) =  \sum_{m=1}^{\infty}{\cos(\pi m \tau) \over \sqrt{m}} {\cosh(\, |1-x| \gamma\sqrt{m}\, )\over  \sinh (\gamma \sqrt{m})} \, ,
\label{lambda1_1}
\ee
\be
\lambda_0 =  c_0 - \sum_{m=1}^{\infty}{1 \over \sqrt{m}} \left[  \coth (\gamma \sqrt{m})  -1\right] \, ,
\label{lambda1_0}
\ee
\be
 c_0 = \!\!\!\ \lim_{m_*\to \infty} \left[ 2 \sqrt{m_*+1/2} \, -\,  \sum_{m=1}^{m_*} {1 \over \sqrt{m}} \right]  \approx 1.460 \, .
\label{c0}
\ee
(The term $1/2$ under the square root dramatically enhances the convergence.)

%Of special interest---in particular, in the experimental context---is
%the single-particle density matrix $\rho(x)=G(x,0)$ at
%nonzero temperature and finite system size, that, based on above analysis,
%is given by

Of particular experimental interest is the single-particle density matrix $\rho(x)=G(x,0)$ at nonzero temperature and finite system size given by
\be \rho(x)= n_0 \Phi(x)
 \exp \left\{ - {L x(2-x) \over 8n_s\beta } + {\sqrt{D}\, [c_0 +
    \lambda_{\rho}(x) ] \over n_s \sqrt{2\pi \beta}} \right\} ,
\label{rho}
\ee
\be
\lambda_{\rho}(x)\! = \!\! \sum_{m=1}^{\infty} {1\over\sqrt{m}} \left[ 1 - {\cosh(\gamma \sqrt{m}) - \cosh(|1\! - \!x| \gamma\sqrt{m})\over  \sinh (\gamma \sqrt{m})} \right] \! .
\label{lambda_rho}
\ee
The dependence of $\rho$ on $\beta$ and $L$ cannot be
reduced to a single scaling combination of these
parameters, meaning that both can be used
independently for probing and verifying the universal off-diagonal
correlations.  The divergence of $\lambda_1(x,\tau)$ and $\lambda_\rho(x)$ at $x, \tau \to 0$  signals their sensitivity to UV cutoff
[formally taken to infinity in (\ref{lambda1_1}) and (\ref{lambda_rho})]; this divergence is eliminated inside $\tilde{C}(x,\tau)$ by the UV-cutoff-dependent condensate density, as is clear from (\ref{phase_corr}).

In Fig.~\ref{G}, we show the result of worm-algorithm
quantum Monte Carlo simulations \cite{Worm} of the minimal model
with relatively large system size in the $y$-direction, but
smaller length $L$ in the iTQF direction. The temperature
$ T \sim L^{-1}  \gg t/L^2$  should be considered high given that the low-frequency
iTQF dynamics is characterized by $\omega \sim Dk^2$.
Since $n_s$  is computed with high accuracy from statistics of path-integral
winding number fluctuations, two parameters, $n_0$ and $D$,  describe all Green's function data in the space-time domain.
Clearly, $n_0$ is responsible only for the overall signal amplitude at
large $x$ and $\tau$. Thus $D$ is solely responsible for the shape of
the curves. Apart from providing perfect description of all data in the asymptotic limit,
the theory works remarkably well down to the lattice constant distance
and $\tau t \lesssim 1$, see left panel in Fig.~\ref{G}.

The central panel in Fig.~\ref{G} demonstrates that non-zero temperature effects
are under precise theoretical control. In this case, simulation data for lower temperature
(larger $\beta$) are reproduced by simply taking parameters deduced from the higher temperature
fits. The right panel in Fig.~\ref{G} is more relevant to the possible experimental observation
of the iTQF state: while it is relatively standard to recover the density matrix from the measured
momentum distribution, there is no easy direct access to the Green's function in imaginary time.
However, if several density matrixes at different temperatures are measured, their joint fit
will provide access not only to the iTQF ground state parameters, but also constitute an accurate
thermometry protocol.  Moreover, the difference between the solid and dashed lines for
$\beta t =16$ case indicates that the density matrix data at elevated temperature are
sensitive enough to resolve the persistent currents contribution  
and, in particular, can be used to estimate $n_s$ directly from $\Phi(x) - 1 \approx  -2 e^{- \beta E_1}  [1-\cos (2\pi x/L) ]  $.

%%%%%%%%%%%%%%%%%%%%%%%%%%%%%%%%%%%%%%%%%%%%%%%%%%
%%%%%%%%%%%%%%%%%%%%%%%%%%%%%%%%%%%%%%%%%%%%%%%%%%%%%%%

In order to further show the universality of the theory, 
we have also simulated the microscopic model of Ref.~\cite{Yan}, 
\begin{equation}
H = - \sum_i a_i^{\dagger} a_{i+1}   -  \sum_{i,j} b_{i,j}^{\dagger} b_{i, j+1} - t_{\perp} \sum_i a_i^{\dagger} b_{i,0} + {\rm H.c.} ,
\label{eq:Zi}
\end{equation}
where the $a_i$ bosons are moving along the $x$-axis as in Fig.~\ref{iTQF_model}, and the $b$ bosons---along the $y$-axis,
with the hopping amplitude $t_{\perp}$ between the $a$ and $b$ bosons. 
Despite microscopic differences from the model in Fig.~\ref{iTQF_model}, model (\ref{eq:Zi}) exhibits the same low-energy universal iTQF phenomenology. We have demonstrated this for the system with $L = \beta = 64$, $L_y = 128$, and $t_{\perp}= 1$. The correlation functions and the corresponding fits are presented in Fig.~\ref{fig:Zi}: Although the fit has been performed for the equal time single-particle density matrix, the two fit parameters fully specify the other Green functions as well~\cite{footnote}. 

{\it TQF case.} Straightforward integration over $\omega$ results in
a Gaussian integral over $k$,  leading to the transparent final answer:
\be
C (x,\tau)  \,=\, {\sqrt{D} \, e^{-{x^2 \over 4 D |\tau |}} \over 4 n_s \sqrt{\pi |\tau |}}    \qquad \mbox{(TQF)} \, .
\label{Lambda_TQF}
\ee
We do not elaborate further on the non-zero temperature and finite system size
expressions because they can be obtained by following the procedure described above for
iTQF identically.
%%%%%%%%%%%%%%%%%%%%%%%%%

{\it Summary and conclusion.}  Motivated by a number of recently
proposed physical realizations, we demonstrated how an effective field theory
can be used to make detailed predictions for space-time correlations for
finite-size systems at nonzero-temperatures for two types of
transverse quantum fluids, which constitute a new class of quasi-one-dimensional
quantum fluids that even in 1D exhibit an off-diagonal (superfluid)
long-range-ordered ground states. Perfect agreement with quantum Monte
Carlo simulations allows us to state that all effective field theory parameters,
and even system temperature, can be reliably extracted from the
one-particle density matrix measurements in finite-size systems, e.g. in
cold-atom-engineered TQFs. We propose that such benchmarking experiments
be used for measuring other TQF properties such as real-time dynamics,
entanglement, out-of-time-order correlators (OTOC), full-counting distribution functions, to name a
few, because computing them is difficult or even impossible by existing
analytical treatments or numerical simulations.

LR's research was supported by the Simons Investigator Award from the
Simons Foundation. LR thanks The Kavli Institute for Theoretical
Physics for hospitality while this manuscript was in preparation,
during Quantum Crystals and Quantum Magnetism workshops, and support from
the National Science Foundation under Grant No. NSF PHY-1748958,
PHY-2309135. LP acknowledges funding by the Deutsche Forschungsgemeinschaft (DFG, German Research Foundation) under Germany's Excellence Strategy -- EXC-2111 -- 390814868.
AK, BS, and NP acknowledge support from the National Science Foundation under Grants DMR-2032136 and DMR-2032077. Some of the quantum Monte Carlo codes~\cite{Worm_scipost} make use of the ALPSCore libraries~\cite{ALPSCore1, ALPSCore2}.

\end{document}